\RequirePackage{ifpdf}
\ifpdf 
\documentclass[pdftex]{sigma}
\else
\documentclass{sigma}
\fi

\def\erfc{{\rm erfc}\,}
\def\erf{{\rm erf}\,}

\begin{document}

\renewcommand{\PaperNumber}{020}

\FirstPageHeading

\ShortArticleName{Exact Propagators for Soliton Potentials}

\ArticleName{Exact Propagators for Soliton Potentials}

\Author{Andrey M. PUPASOV and Boris F. SAMSONOV}
\AuthorNameForHeading{A.M. Pupasov and B.F. Samsonov}
\Address{Department of Physics, Tomsk State University, 36 Lenin
Ave., 634050 Tomsk, Russia}
\Email{\href{mailto:pupasov@phys.tsu.ru}{pupasov@phys.tsu.ru},
\href{mailto:samsonov@phys.tsu.ru}{samsonov@phys.tsu.ru}}

\ArticleDates{Received October 01, 2005, in final form November
21, 2005; Published online November 24, 2005}

\Abstract{Using the method of Darboux transformations (or
equivalently supersymmetric quantum mechanics) we obtain an
explicit expression for the propagator for the one-dimensional
Schr\"odinger equation with a multi-soliton potential.}


\Classification{81Q05; 81Q60}

\section{Introduction}

It is nowadays incontestable that ideas of supersymmetry play an
important role in modern theoretical and mathematical physics.
 Supersymmetric quantum mechanics (SUSY QM), introduced
 by Witten \cite{vitten} as a toy model in
 supersymmetric quantum field theory,
is a very efficient tool for studying different properties of
non-relativistic quantum systems. In particular, as it was
recently shown \cite{my} it may become an essential ingredient of
complex quantum mechanics which is currently under development
since it may ``cure'' such a ``disease''  of non-Hermitian
Hamiltonians as their non-diagonalizability, and can remove
spectral singularities from the conti\-nuous part of the spectrum.
On the other hand, it is well known  \cite{andrianov} that SUSY QM
is basically equivalent to the method of Darboux transformations
\cite{darboux} well known in connection with soliton
theory~\cite{Matveev}. We mean by soliton potentials such
potentials of the one-dimensional stationary Schr\"odinger
equation from which soliton solutions of the Korteweg--de~Vries
equation may be constructed \cite{Matveev}
 and which are SUSY partners of the zero potential.
They find an application in quantum field theory  for describing
processes where solitons may play an essential role \cite{Raj}.

It is not an exaggeration to say that everything in nature
undergoes time-evolution. Therefore, one of the crucial questions
of any physical theory is to describe how a physical phenomenon
evolves with time. In quantum mechanics time dependence of the
wave function may be described with the help of the propagator
which is nothing but the evolution operator in coordinate
representation. As far as we know, the first attempt to find the
propagator for a multi-soliton potential is due to Jauslin
\cite{jauslin}. His method is based on an integral formula which
relates solutions of two Schr\"odinger equations whose Hamiltonians
are SUSY partners. Unfortunately, this approach when applied to
the Schr\"odinger equation  leads to divergent integrals.
Therefore the author found the propagator for the
 heat equation with the one-soliton potential.
 The Schr\"odinger equation may be considered as
 the heat equation with the imaginary time.
 In this respect the
 following question arises: whether or not the Jauslin's  result
 after the replacement $t\to it$
 gives the propagator for the one-soliton potential? We want to
 stress that the answer to this question is not trivial since such
 a replacement at the level of the Jauslin's integral transformation
 leads to divergent integrals. Our analysis shows that the answer
 to this question is positive.

We would like to note that the problem of finding propagators is
more difficult than solving the stationary Schr\"odinger equation
for a time-independent Hamiltonian, since knowledge of the
propagator allows us to solve the Cauchy problem for the
non-stationary equation with the same Hamiltonian, but with an
arbitrary initial condition. This is similar to the Green function
of the stationary equation; knowing it allows us to write down a
solution of the inhomogeneous equation with an arbitrary
inhomogeneity. Recently a method has been proposed for finding the
Green function for a SUSY partner Hamiltonian
\cite{samsonov&sukumar&pupasov}.
 In
this paper, using a particular example of soliton potentials, we
show that the method of SUSY QM is very helpful for finding exact
propagators for Hamiltonians related by SUSY (or equivalently
Darboux) transformations. We would like to stress that although
many aspects of multi-soliton potentials are well-studied in the
literature a closed form of the propagator as far as we know has
not yet been published.

To make the paper self-contained, first we briefly review
 the method we are using and give a~general expression for the
 propagator (Section 2 and the beginning of Section 3),
 and then apply it for finding the
 propagator for the multi-soliton potential (Section 3). Finally, we
 apply our general result to the case of
 one- and two-soliton
 potentials (Section 4) and  suggest a formula for $N$-soliton potential.

\section{Darboux transformations and soliton potentials }

In this section we  review briefly the method of Darboux (SUSY)
transformations \cite{Matveev,BS,SUSY} Also we construct a
differential (i.e.\ Darboux) transformation operator for obtaining
solutions of the Schr\"odinger equation with a soliton potential
from solutions of the free particle equation we need in the
following sections.

  Consider two
one-dimensional Schr\"{o}dinger equations
\begin{gather}\label{1} (h_0-E)\psi_E(x)=0,\qquad
h_0=-\partial_x^2+V_0(x),
\\
\label{2} (h_1-E)\varphi_E(x)=0,\qquad h_1=-\partial_x^2+V_1(x).
\end{gather}
Suppose one knows solutions to equation (\ref{1}). Then solutions
to equation (\ref{2}) can be found by acting with the {\it
transformation operator} (we denote it as $L$) on solutions to
equation (\ref{1}),
 $\varphi_E=L\psi_E$.
 The main relation defining $L$
is the intertwining relation
 \begin{gather}\label{inter}
 Lh_0=h_1L.
 \end{gather}

An essential point of the method is the choice of the operator
$L$. If we restrict $L$ to be a {\it differential operator} it
becomes {\it Darboux transformation operator} (see
e.g.~\cite{BS}). In this case the potential $V_1$ cannot be
arbitrary, and should be found together with the transformation
operator from the intertwining relation~(\ref{inter}).
 If $L$ is a first
order differential operator, the result is well-known
 (see e.g.~\cite{Matveev,BS,SUSY}):
\[
 L=-\partial_x +w(x),\qquad
 V_1=V_0-2w'(x),
\]
  where
\[ w(x)=(\ln u)_x, \qquad
h_0u=\alpha u.
\]
Function $u=u(x)$ and parameter $\alpha$ are called {\it
transformation function}
 and {\it factorization constant} respectively.

Since the procedure is independent on the initial Hamiltonian, it
can be repeated as many times as one desires.
 So, we can take $h_1$ as the initial Hamiltonian for the
next transformation step to get $h_2$ and so on. In this way one
arrives at chains of transformations. It is remarkable that the
resulting action of a chain may be expressed in terms of solutions
of the initial equation only. If all factorization constants are
different from each other, one obtains Crum--Krein~\cite{CK}
formulas
\begin{gather}\label{fiE}
\varphi_E=L\psi_E=W(u_1,u_2,\ldots,u_N,\psi_E)/W(u_1,u_2,\ldots,u_N),
\\
 \nonumber
V_N=V_0-2 \left[\log W(u_1,u_2,\ldots,u_N)\right]''.
\end{gather}
Here and in the following the symbol $W$ denotes a Wronskian, and
$u_j=u_j(x)$ is an eigenfunction of the initial Hamiltonian
(transformation function)
\[
h_0u_j(x)=\alpha_ju_j(x),\qquad j=1,2,\ldots,N.
 \]

To get soliton potentials one starts with the zero initial
potential, $V_0(x)=0$, and uses the following set of $N$ (which is
supposed to be even) transformation functions
\cite{Matveev,BS,Sukumar}
\begin{gather}\label{fpfs}
u_{2j-1}(x)=\cosh(a_{2j-1}x+b_{2j-1}),\\
 u_{2j}(x)=\sinh(a_{2j}x+b_{2j}),\qquad j=1,2,\ldots, N/2.
\end{gather}
They are solutions to the Schr\"odinger equation with the zero
potential corresponding to eigenvalues $E_j=-a_j^2<0$, which are
just the points of the discrete spectrum of
$h_N=-\partial_x^2+V_N(x)$.
 An orthonormal set of its
discrete spectrum eigenfunctions is given by
\begin{gather}\label{ortef}
\varphi_n(x)=\left(\frac{a_n}{2}\prod_{j=1(j\neq
n)}^{N}|a_n^2-a_j^2|\right)^{1/2}
\frac{W^{(n)}(u_1,u_2,\ldots,u_{n-1},u_{n+1},\ldots,u_N)}
{W(u_1,u_2,\ldots,u_N)}.
\end{gather}
Here $W^{(n)}(u_1,u_2,\ldots,u_{n-1},u_{n+1},\ldots,u_N)$ is the
Wronskian of
 order $N-1$ obtained from the Wronskian
$W(u_1,u_2,\ldots,u_N)$ by dropping the function $u_n$.

We also need the continuous spectrum eigenfunctions of $h_N$,
which should be found by acting with the operator $L$ (\ref{fiE})
on plane waves $\psi_k(x)=1/\sqrt{2\pi}\exp(-ikx)$, $k\in {\mathbb R}$
\begin{gather}\label{fik}
\varphi_k(x)=\frac{1}{\sqrt{(k^2+a_1^2)(k^2+a_2^2)\cdots(k^2+a_N^2)}}
L\psi_k(x),
\\
E=k^2,\qquad \alpha_k=-a_k^2,\qquad k=1,\ldots,N.\nonumber
\end{gather}
The set of functions $\{\varphi_n(x)$, $n=1,\ldots,N\}$ and
$\{\varphi_k(x)$, $k\in{\mathbb R}\}$ forms a complete and orthonormal
set in the Hilbert space of square integrable functions on the
whole real line.

It is interesting to note that for particular values of the
parameters $a_j$ a multi-soliton potential may have a shape of a
multi-well potential thus presenting an example of a multi-well
exactly solvable potential.

\section{Propagator for a multi-soliton potential}

We use the definition of the propagator $K(x,y;t',t'')$ of the
Schr\"{o}dinger equation
 as the coordinate representation of the evolution operator
(see e.g.~\cite{Tanudji}). If one knows solutions of the
non-stationary Schr\"odinger equation with a time-independent
potential, which form a complete and orthonormal
 set
(for continuous subset, if present, normalization is understood in
the sense of generalized functions)
  in the Hilbert
space of square integrable functions on the whole real line,
 then the propagator is given
by
\[ K(x,y;t',t'')=\theta(t'-t'')\left[
\sum_{n=0}^{N}\psi_n^*(x,t')\psi_n(y,t'')+ \int_{-\infty}^\infty
dk\psi_k^*(x,t')\psi_k(y,t'')\right].
\]
Here $\{ \psi_n(x,t)\}$ is the discrete part of the basis and
$\{\psi_k(x,t),\ k\in {\mathbb R}\}$ is  the continuous one. Everywhere
we shall assume $t'>t''$ and drop the step function
$\theta(t'-t'')$.

According to this formula, the propagator has two contributions
\begin{gather}\label{Kdc}
 K(x,y;t',t'')=K_d(x,y;t',t'')+K_c(x,y;t',t'').
 \end{gather}
The first term at the right hand side of formula (\ref{Kdc}),
$K_d$, corresponds to the discrete part of the basis, and the
second one, $K_c$, proceeds from the continuous part. Our crucial
observation is that the method of Darboux transformations gives
both the discrete and the continuous parts of the basis for the
transformed equation if a complete set of eigenfunctions for the
initial Hamiltonian is known.

For soliton potentials the discrete part is given by (\ref{ortef})
and the continuous one is given by (\ref{fik}). Therefore
\begin{gather}\label{cd}
K_{dN}=\sum_{n=0}^{N}\varphi_n^*(x,t')\varphi_n(y,t''),\qquad
K_{cN}=\int_{-\infty}^\infty dk\varphi_k^*(x,t')\varphi_k(y,t''),
\end{gather}
where the additional subscript $N$ labels the order of the Darboux
transformation, which in our case coincides with the number of
solitons.

We compute first the value $K_{cN}$. After inserting
$\varphi_k(x)$ from (\ref{fik})
 into
\eqref{cd} we interchange derivatives with the integrals over the
momentum $k$.
 This allows us to
present the contribution from the continuous spectrum as an action
of the transformation operator $L$ on an integral
\begin{gather}
\nonumber
 K_{cN}(x,y;t',t'')   = L_xL_y\int_{-\infty}^\infty
dk\frac{\psi_k^*(x,t')\psi_k(y,t'')}{(k^2+a_1^2)(k^2+a_2^2)\cdots(k
^2+a_N^2)}\\
\phantom{K_{cN}(x,y;t',t'')}{} =
 \frac{1}{2\pi}L_xL_y\int_{-\infty}^\infty
dk\frac{e^{ik(x-y)+ik^2(t'-
t'')}}{(k^2+a_1^2)(k^2+a_2^2)\cdots(k^2+a_N^2)}. \label{intKc}
 \end{gather}
Here $L_x$ is given by (\ref{fiE}) and $L_y$ is obtained from
$L_x$ by the replacement $x\to y$. The integral in (\ref{intKc})
is calculated in Appendix. Thus, the contribution from the
continuous part of the basis has the form:
\begin{gather}\label{kc}
K_{cN}(x,y,t',t'')
=\frac{1}{2\pi}\sqrt{\frac{it}{2}}L_xL_y\sum_{n=1}^N\left(\prod_{m
\neq n}^N\frac{1}{\alpha_n-\alpha_m}\right)
I\left(a_n\sqrt{\frac{it}{2}},\sqrt{\frac{i}{8t}}(x-y)\right).
\end{gather}
Here $t=t'-t''$ and
\begin{gather}\label{I}
 I(a,x)
 =\int_{-\infty}^\infty dp
\frac{{\rm e}^{-2p^2+4px}}{p^2+a^2} = \frac{\pi}{a}{\rm e}^{2a^2}
\mbox{Re} [{\rm e}^{4ixa}\erfc(a \sqrt{2}+ix\sqrt{2})].
\end{gather}

To compute $K_{dN}$, we simply replace in  (\ref{cd}) the discrete
basis eigenfunctions $\varphi_i$ according to~\eqref{ortef} and
take into account their time dependence, which yields
 \begin{gather}\label{kd}
K_{dN}=\sum_{n=0}^{N}\left(\frac{a_n}{2}\prod_{j=1(j\neq
n)}^{N}|a_n^2-
a_j^2|\right)\frac{W^{(n)}(x)W^{(n)}(y)}{W(x)W(y)}{\rm
e}^{ia_n^2t},\qquad t=t'-t''.
\end{gather}

Thus, we see that the
 propagator for the multi-soliton potential is expressed in terms of
the error function, derivative of the error function and solutions
corresponding to the discrete part of the basis. In the next
Section, we suggest a simpler formula for the propagator.

\section{Particular cases}

Let us consider how expressions (\ref{kc}) and (\ref{kd}) lead to
the propagators for the one- and two-soliton potentials.

The one-soliton potential
\[
V(x)=-2a^2/\cosh^2(ax)
\]
may be obtained from the zero potential with help of the first
order Darboux transformation with the transformation function
$u(x)=\cosh(ax)$. The only discrete spectrum eigenfunction is
$\varphi_0(x)=\big(a/\sqrt{2}\big)/\cosh(ax)$.

After some simple calculations we obtain from (\ref{Kdc}),
(\ref{kc})--(\ref{kd}) the propagator for the Schr\"{o}\-din\-ger
equation with the one-soliton potential
\begin{gather}\label{ospp}
K_1(x,y;t'',t')=\frac{1}{\sqrt{4\pi it}} {\rm{e}}^{
\frac{i(x-y)^2}{4t}}+\frac{a
{\rm{e}}^{ia^2t}}{4\cosh(ax)\cosh(ay)}\left[\erf_{+}(a)
+\erf_{-}(a)\right], \\
 \nonumber t=t'-t''.
\end{gather}
Here $\erf_{\pm}(a)$ is the error function taken at a special
value of the argument
\[
\erf_{\pm}(a)=\erf
\left[a\sqrt{it}\pm\frac{i\sqrt{i}}{2\sqrt{t}}(x-y)\right].
\]

Now two observations are in order. (i) We state that up to the
replacement $t\to it$ our result is in the perfect agreement with
that of Jauslin \cite{jauslin}. This fact justifies the Jauslin's
procedure of regularization of corresponding divergent integral.
And (ii) we see
 that the propagator (\ref{ospp}) has two contributions.
 The first term coincides with the
propagator for the free particle and, hence, the second one
describes just the one-soliton perturbation of the free particle
at the level of the propagator.

 In case of the two-soliton potential
the method of Jauslin becomes very involved, since it is not clear
how corresponding integral may be calculated. In contrast, our
formulas (\ref{kc}) and (\ref{kd}) give an explicit expression for
the propagator. To use formula (\ref{kd}), we first calculate
necessary Wronskians with the transformation functions
(\ref{fpfs}), where for simplicity we choose $b_1=b_2=0$, i.e.\ $
u_1(x)=\cosh(a_1x)$, $u_2(x)=\sinh(a_2x)$ and then after some
algebra we get the propagator for the two-soliton potential
\begin{gather}
K_2(x,y;t'',t')=  \frac{1}{\sqrt{4\pi it}} {\rm{e}}^{
\frac{i(x-y)^2}{4t}}\nonumber\\
\phantom{K_2(x,y;t'',t')=}{}
+\frac{a_1(a_2^2-a_1^2)\sinh(a_2x)\sinh(a_2y)
{\rm{e}}^{ia_1^2t}}{4W(x)W(y)}\left[\erf_{+}(a_1)
+\erf_{-}(a_1)\right]
\nonumber\\
\phantom{K_2(x,y;t'',t')=}{}
+\frac{a_2(a_2^2-a_1^2)\cosh(a_2x)\cosh(a_2y)
{\rm{e}}^{ia_2^2t}}{4W(x)W(y)}\left[\erf_{+}(a_2)
+\erf_{-}(a_2)\right],\label{tspp}\\
t=t'-t''.\nonumber
\end{gather}

Once again we see that the propagator for the
 two-soliton potential
  has a structure similar to that of the one-soliton potential.
  Both the propagator (\ref{ospp}) and (\ref{tspp}) have
  two contributions:
  the exponential part which is the free
particle propagator
 and the part responsible for the soliton perturbation of the zero
 potential.

This result suggests us the form of the propagator of the general $N$-soliton
potential as a~sum of two contributions, namely the free particle
propagator in the form of the exponential function and an
$N$-soliton perturbation:
\begin{gather}\nonumber
K_N(x,y;t'',t')=  \frac{1}{\sqrt{4\pi it}} {\rm{e}}^{
\frac{i(x-y)^2}{4t}}+\sum_{n=0}^{N}\left(\frac{a_n}{4}\prod_{j=1(j\neq
n)}^{N}|a_n^2-
a_j^2|\right)\nonumber\\
\phantom{K_N(x,y;t'',t')=}{}
\times\frac{W^{(n)}(x)W^{(n)}(y)}{W(x)W(y)}{\rm{e}}^{ia_n^2 t}
\left[\erf_{+}(a_n) +\erf_{-}(a_k)\right],\qquad t=t'-t''.\label{nspp}
 \end{gather}
We have checked this formula  for $N=3,4$ by the direct
substitution into the Schr\"odinger equation.

\section{Conclusion}

In this paper using the known approach based on supersymmetric
quantum mechanics (or equi\-va\-lently, the method of Darboux
transformation) we construct propagators for multi-soliton
potentials. In particular, we present explicit expressions for the
case of one- and two-soliton potentials. While comparing these
results with the ones previously published by Jauslin
\cite{jauslin}, we conclude that the transformation
$t\leftrightarrow it$ (Whick rotation) is justified for
regularization of divergent integrals of a special type. 
Finally, we suggest an explicit expression for the propagator for the
general multi-soliton potential.

\subsection*{Acknowledgements}

This work is partially supported by the President Grant of Russia
1743.2003.2.

\appendix

\section{Appendix}

Here we calculate the integral
\[
\int_{-\infty}^\infty dk\frac{{\rm e}^{ikx+ik^2t}}%
{(k^2+a_1^2)(k^2+a_2^2)\cdots(k^2+a_N^2)}.
\]
The identity
\[
\prod_{n=1}^{N}\frac{1}{k^2+a_n^2}=\sum_{n=1}^N
\left(\prod_{j=1,j\neq
n}^N\frac{1}{a_j^2-a_n^2}\right)\frac{1}{k^2+a_n^2}
\]
reduces it to a sum of simpler integrals of the form:
\[
\int_{-\infty}^\infty dk\frac{{\rm e}^{ikx+ik^2t}}%
{k^2+a^2}=\frac{1}{2ia}\int_{-\infty}^\infty dk{\rm e}^{i
kx+ik^2t}\left(\frac{1}{k-ia}-\frac{1}{k+ia}\right).
\]
One can find the value of the integral at the right hand side of
this equation in \cite{Manko}
\[
\int_{-\infty}^\infty dp\frac{{\rm
e}^{-2p^2+4px}}{p+i\alpha}=-i\pi {\rm
e}^{2\alpha(\alpha-2ix)}\erfc\big[\sqrt{2}(\alpha-ix)\big]
\]
which leads just to formula (\ref{I}) for the function
\[
 I(a,x)=\int_{-\infty}^\infty dp
\frac{{\rm e}^{-2p^2+4px}}{p^2+a^2}.
\]

\LastPageEnding

\end{document}